\documentclass[12pt,preprint]{aastex}

\begin{document}

\shorttitle{Wave Properties of Coronal Bright Fronts}
\shortauthors{Long et al.}

\title{The Wave Properties of Coronal Bright Fronts Observed Using SDO/AIA}

\author{David M. Long\altaffilmark{1} Edward E. DeLuca}
\affil{Harvard-Smithsonian Center for Astrophysics, 60 Garden Street, Cambridge, MA 02138, USA}
\email{longda@tcd.ie}

\author{Peter T. Gallagher}
\affil{Astrophysics Research Group, School of Physics, Trinity College Dublin, Dublin 2, Ireland}

\altaffiltext{1}{Astrophysics Research Group, School of Physics, Trinity College Dublin, Dublin 2, Ireland}

\begin{abstract}
Coronal bright fronts (CBFs) are large scale wavefronts that propagate though the solar corona at hundreds of kilometers per second. While their kinematics have been studied in detail, many questions remain regarding the temporal evolution of their amplitude and pulse width. Here, contemporaneous high cadence, multi-thermal observations of the solar corona from the \emph{Solar Dynamic Observatory} (\emph{SDO}) and \emph{Solar TErrestrial RElations Observatory} (\emph{STEREO}) spacecraft are used to determine the kinematics and expansion rate of a CBF wavefront observed on 2010~August~14. The CBF was found to have a lower initial velocity with weaker deceleration in \emph{STEREO} observations compared to \emph{SDO} ($\sim$340~km~s$^{-1}$ and $-72$~m~s$^{-2}$ as opposed to $\sim$410~km~s$^{-1}$ and $-279$~m~s$^{-2}$). The CBF kinematics from \emph{SDO} were found to be highly passband-dependent, with an initial velocity ranging from $379\pm12$~km~s$^{-1}$ to $460\pm28$~km~s$^{-1}$ and acceleration ranging from $-128\pm28$~m~s$^{-2}$ to $-431\pm86$~m~s$^{-2}$ in the 335~\AA\ and 304~\AA\ passbands respectively. These kinematics were used to estimate a quiet coronal magnetic field strength range of $\sim$1--2~G. Significant pulse broadening was also observed, with expansion rates of $\sim$130~km~s$^{-1}$ (\emph{STEREO}) and $\sim$220~km~s$^{-1}$ (\emph{SDO}). By treating the CBF as a linear superposition of sinusoidal waves within a Gaussian envelope, the resulting dispersion rate of the pulse was found to be $\sim$8--13~Mm$^2$~s$^{-1}$. These results are indicative of a fast-mode magnetoacoustic wave pulse propagating through an inhomogeneous medium.
\end{abstract}


\section{Introduction}
\label{sect:introduction}

Coronal bright fronts (CBFs; commonly called ``EIT Waves'') in Extreme UltraViolet (EUV) observations of the low solar corona were first noted by \citet{Moses:1997vn} and characterised by \citet{Thompson:1998ab}. They have since been studied extensively using EUV observations from the \emph{SOHO} \citep{Thompson:1999cd}, \emph{TRACE} \citep{Wills-Davey:1999ab}, \emph{STEREO} \citep{Long:2008eu} and most recently \emph{SDO} \citep{Liu:2010ab} spacecraft. CBFs are usually observed as diffuse bright fronts propagating isotropically when unimpeded at typical velocities of 200--400~km~s$^{-1}$ across the solar disk \citep{Thompson:2009yq}. They are best observed using the 195~\AA\ passband, at a temperature of 1--2~MK and height of $\sim$70--90~Mm above the photosphere \citep{patsourakos2009, kienreich2009}. However, they have also been observed in other passbands, including 171~\AA\ \citep{Wills-Davey:1999ab}, 284~\AA\ \citep{Zhukov:2004kh}, 304~\AA\ \citep{Long:2008eu} and the 94, 131, 211, and 335~\AA\ passbands monitored by the Atmospheric Imaging Assembly (AIA) onboard \emph{SDO} \citep{Liu:2010ab}.

Despite more than 15 years of study using observations from ground-based \citep{Gilbert:2004ab,Chen:2009ab} and space-based \citep{Attrill:2009ab,Chen:2010ab} instruments, CBFs remain an enigma with many competing theories attempting to explain this phenomenon. They have been alternatively interpreted as magnetohydrodynamic waves \citep{Wang:2000tg,Warmuth:2004ab,Wang:2009ab,Schmidt:2010ab}, solitons \citep{Wills-Davey:2007oa} and in terms of magnetic field restructuring during the eruption of an associated CME \citep{Chen:2002rw, Attrill:2007vn, Delannee:2008uq}. For a detailed description of CBFs, see the recent reviews by \citet{Gallagher:2010ab} and \citet{Wills-Davey:2010ab}.

Traditional analysis of CBFs has produced kinematics that are inconsistent with MHD wave theory, implying a pseudo-wave interpretation. However, recent observations of decelerating CBFs combined with the effects of low observing cadence \citep{Long:2008eu,Ma:2009ab} suggest that this may not be the case. There have also been indications of CBF dispersion with propagation \citep{Warmuth:2004ab,Long:2011ab}, although this has been difficult to quantify. While these properties are inconsistent with ideal MHD wave theory, they have been shown in simulations by \citet{Murawski:2001ab} and \citet{Nakariakov:2005ab} to be a natural result of propagation through an inhomogeneous medium.

\emph{SDO}/AIA observes the Sun continuously at a cadence of $\sim$12~s in seven EUV passbands, an improvement on both \emph{SOHO}/EIT ($\sim$900~s in one of four passbands) and \emph{STEREO}/EUVI ($\sim$75--600~s in four passbands). While this will enable a deeper understanding of the solar corona across a wide range of temperatures, the resulting data volume ($\sim$1.5~TB per day) has necessitated the development of both automated and semi-automated CBF detection and tracking algorithms \citep{podladchikova2005,Wills-Davey:2006ab,Long:2011ab}. Here, the semi-automated CBF algorithm outlined by \citet{Long:2011ab} is applied to \emph{SDO} and \emph{STEREO} observations of the 2010~August~14 CBF and used to determine its physical characteristics.

\section{Observations and Data Analysis}
\label{sect:obs}

The 2010~August~14 CBF event\footnote{Solar Object Locator: SOL2010-08-14T09:38:00L353C79} erupted from NOAA active region (AR) 11093, with an associated coronal mass ejection (CME) and GOES C4.4 flare which started at 09:38~UT. The AR location (N11W65) meant that the on-disk CBF evolution was visible from \emph{STEREO}-A and \emph{SDO} but not \emph{STEREO}-B. When the eruption occurred, \emph{STEREO}/EUVI-A had an observing cadence of 300~s and 600~s in the 195~\AA\ and 304~\AA\ passbands respectively, with the 171~\AA\ and 284~\AA\ passbands both taking synoptic data (i.e.,\ one image every two hours).  \emph{SDO}/AIA was taking observations with 12~s cadence in all seven EUV passbands (94, 131, 171, 193, 211, 304, and 335~\AA) over the same time period. The event is shown in the accompanying movies, with windowed running difference (RD) movies used for the 193 ($I_t - I_{t-4}$; movie\_1.mov) and 304~\AA\ ($I_t - I_{t-10}$; movie\_2.mov) passbands from \emph{SDO} due to the very small relative intensity change between consecutive images. A normal RD movie was used to show the \emph{STEREO} 195~\AA\ passbands ($I_t - I_{t-1}$; movie\_3.mov).

\begin{figure*}[!t]
\centering
   \includegraphics[width=1\textwidth,clip=,trim=0mm 0mm 0mm 0mm]{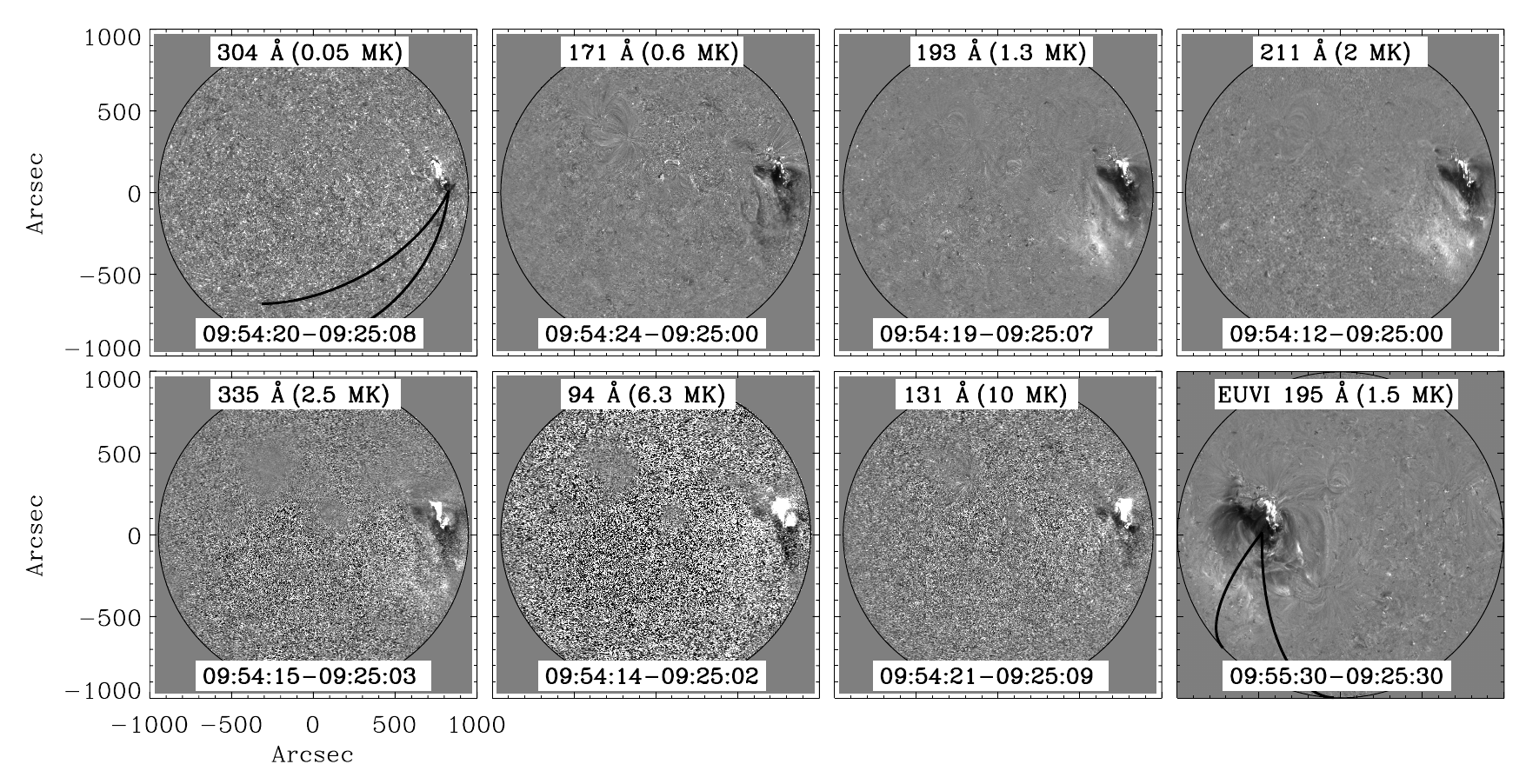}
\caption{PBD images for all \emph{SDO}/AIA passbands and the \emph{STEREO}/EUVI~195~\AA\ passband. Image times used are given on the bottom of each panel. The arc sectors used to identify the pulse are marked in the AIA 304~\AA\ and EUVI 195~\AA\ panels respectively.}
\label{fig:image_panel}
\end{figure*}

The semi-automated detection algorithm used to identify and track the CBF in both EUVI and AIA data works in several steps \citep[see][for more details]{Long:2011ab}. The CBF source location was first defined using the mean centre of ellipses fitted to the first three observations of the CBF in both 193~\AA\ (AIA) and 195~\AA\ (EUVI) data, giving a source unique to both spacecraft (although both sources are comparable when transformed between spacecraft). Percentage base difference (PBD) images \citep{Wills-Davey:1999ab} were used for this analysis, with each image de-rotated to the same time (09:20:30~UT) to compensate for solar rotation and a pre-event time $\sim$09:25:00~UT used to define the base image in each passband (see Figure~\ref{fig:image_panel}). An arc sector was then positioned to allow comparison of both AIA and EUVI observations.%%%, with the resulting intensity profiles used to identify and track the CBF.

The PBD intensity of a given image was averaged across the position angle of the arc sector in annuli of increasing radii and 0.5$^{\circ}$ width on the spherical surface, with the standard deviation giving the associated error. The resulting intensity profile was fitted using a Gaussian function, with the centroid and full width at half maximum (FWHM) giving the pulse position and width respectively. Each parameter has an associated error, quantifying the ability of the algorithm to detect the pulse. Although the source point position and orientation of the arc sector is determined by the user, the actual pulse detection is automated, allowing unbiased identification of the CBF. Once the intensity profiles for each image have been processed and fitted, the CBF is identified as a moving pulse, with any stationary bright features ignored.

\section{Results}
\label{sect:results}

Pulse identification was found to be strongly influenced by passband rather than instrument, with the pulse observed in the 195~\AA\ and 304~\AA\ passbands from \emph{STEREO}/EUVI (although only the 195~\AA\ data was used here due to the low 304~\AA\ cadence). In \emph{SDO}/AIA, the pulse was tracked in four of seven passbands (193, 211, 304, and 335~\AA) with the nature of the 94~\AA\ and 131~\AA\ passbands making identification difficult. Although a slight intensity decrease was visually identified in the 171~\AA\ passband, this could not be tracked using the algorithm.

\subsection{Kinematics}
\label{subsect:kinematics}

The pulse kinematics were determined by measuring the temporal variation in pulse centroid distance from the source point. The bootstrapping technique discussed by \citet{Long:2011ab} was then used to fit a model of the form,
\begin{equation}
r(t) = r_{0} + v_{0}t + \frac{1}{2}at^2
\end{equation}
where $r_{0}$ is the initial distance of the pulse from the source point, $v_{0}$ is the initial velocity and $a$ is the constant acceleration. 

The kinematics of the individual \emph{SDO} passbands are shown in the top-left panel of Figure~\ref{fig:kinematics} to be comparable, although they do tend to separate with propagation. Similarly, the \emph{SDO} 193~\AA\ and \emph{STEREO} 195~\AA\ measurements (top-right panel of Figure~\ref{fig:kinematics}) appear homologous, with a slight positional offset due to the different spacecraft positions.

The kinematics given in Table~\ref{tbl:kinematics} show a lower initial velocity and much weaker acceleration in the 195~\AA\ passband relative to the comparable 193~\AA\ passband. The kinematic estimates from \emph{STEREO} 195~\AA\ are consistent with previous results derived using \emph{SOHO}/EIT and \emph{STEREO}/EUVI, while the higher initial velocity and acceleration from \emph{SDO} 193~\AA\ for the same event suggests a strong influence from the cadence of the observing instrument \citep[cf.][]{Long:2008eu}. The larger uncertainties associated with the kinematics analysis of the 304~\AA\ passband may be explained by the nature of the passband and also by a data gap, which complicated detection of the pulse.

\subsection{Pulse Broadening}
\label{subsect:broadening}

\begin{figure*}[!t]
\centering
   \includegraphics[width=1\textwidth,clip=,trim=0mm 0mm 0mm 0mm]{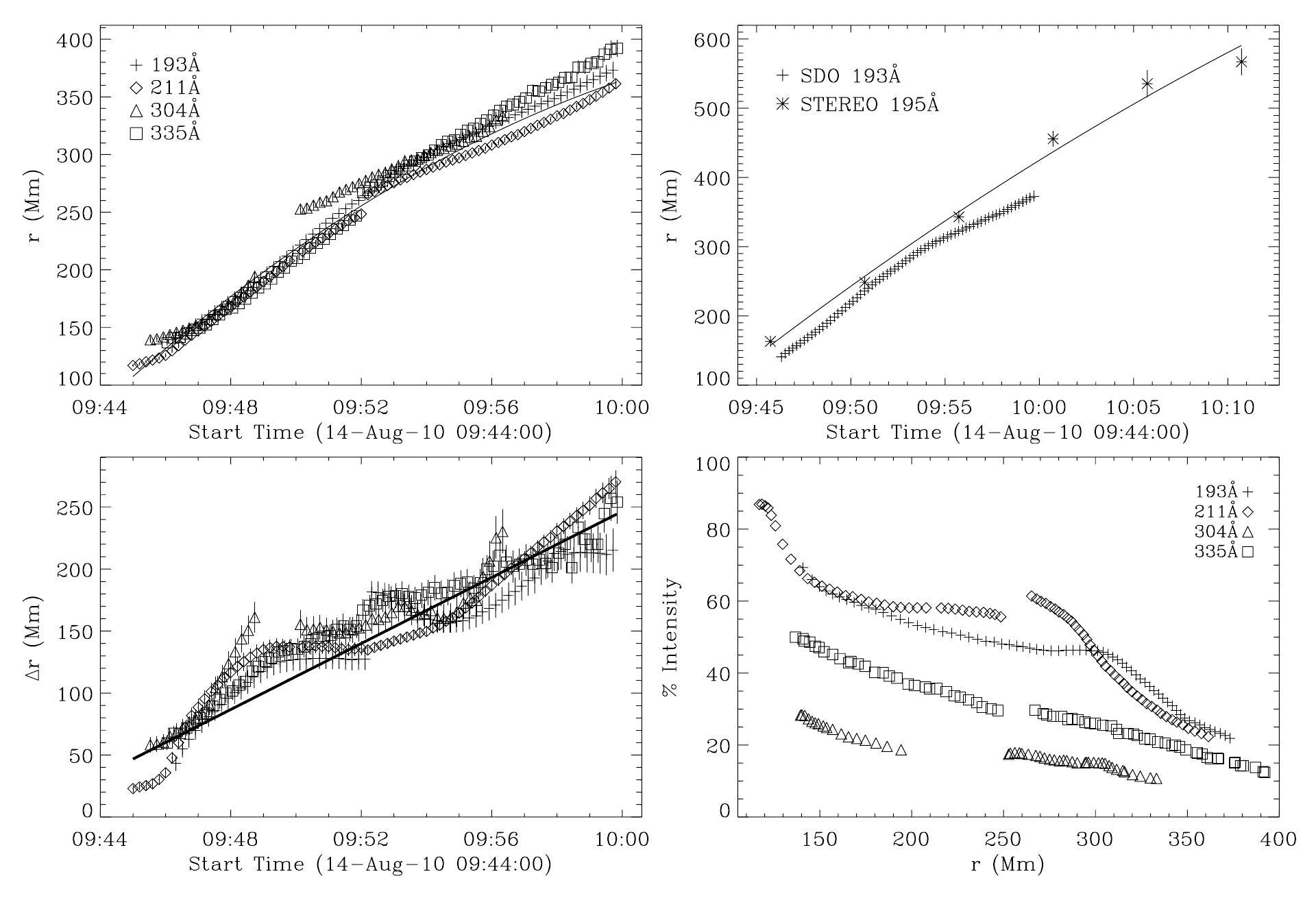}
\caption{\emph{Top left}: Distance-time measurements from AIA 304, 193, 211, and 335~\AA\ passbands for the 2010~August~14 CBF event. \emph{Top right}: AIA 193~\AA\ and EUVI 195~\AA\ distance-time measurements for the same event (line shows the best fit to EUVI measurements). \emph{Bottom left}: Temporal variation in FWHM for AIA passbands with line showing fit to combined \emph{SDO} data; EUVI 195~\AA\ measurements show a similarly increasing trend over a much longer time range. Measurements prior to 09:52~UT have been corrected using a constant offset to remove the effects of a nearby stationary feature. \emph{Bottom right}: Peak PBD pulse intensity variation with distance.}
\label{fig:kinematics}
\end{figure*}

The temporal variation in FWHM was examined for evidence of pulse broadening. The bottom-left panel in Figure~\ref{fig:kinematics} shows that the pulse width changes from $\sim$40~Mm to $\sim$270~Mm over a time period of $\sim$900~s. The data prior to $\sim$09:52~UT has been corrected to remove the effects of a stationary bright feature close behind the CBF. This feature initially exerts a strong influence on the Gaussian fit, but was negated by subtracting a constant offset value for each passband from the FWHM measurements. From $\sim$09:52~UT onward, the CBF is sufficiently far from this feature that the fit to the data is no longer affected. The effects of this bright feature can also be seen in the bottom-right panel in Figure~\ref{fig:kinematics}, which shows the peak \% intensity variation with distance. While the bright feature does initially influence the pulse width and peak intensity variation, a general increase and decrease is apparent for the pulse width and peak intensity respectively.

The dispersion was examined by treating the CBF as a linear superposition of sinusoidal waves within a Gaussian envelope, giving the equation,
\begin{equation}
\psi(r,t) \simeq \textrm{exp}\left(-\frac{(r - v_{g}t)^{2}}{2\sigma^{2}_{r}}\right)\textrm{cos}(k_{0}r - \omega_{0} t)
\end{equation}
where $k_{0}$ is the wavenumber, $\omega_{0}$ is the angular frequency, $\sigma_{r}$ is the characteristic width and $v_{g}$ is the pulse group velocity ($v_{g} = d\omega/dk$). The pulse extends in Fourier space from $k_{0} - \Delta k/2$ to $k_{0} + \Delta k/2$ ($\Delta k \sim \sigma_k$ where $\sigma_k = 1/\sigma_r$), so that the velocity varies from $v_{g}(k_{0} - \Delta k/2)$ to $v_{g}(k_{0} + \Delta k/2)$ across the pulse. The pulse therefore broadens with propagation, with a spatial extent (FWHM) defined as,
\begin{equation}
\Delta r(t) = \Delta r_{0} + \left[v_{g}\left(k_{0} + \frac{\Delta k}{2}\right) - v_{g}\left(k_{0} - \frac{\Delta k}{2}\right)\right]t
\end{equation}
where $\Delta r_{0}$ is the initial pulse width. This can be rewritten in terms of the change in group velocity $v_{g}$ as,
\begin{equation}
\Delta r(t) \sim \Delta r_{0} + \frac{dv_{g}(k_{0})}{dk}\Delta kt.
\end{equation}
As the group velocity $v_{g} = d\omega/dk$, the width of a dispersive pulse at any time $t$ is given by,
\begin{equation}\label{eqn;disp}
\Delta r(t) = \Delta r_{0} + \frac{d^2 \omega(k_{0})}{dk^{2}}\frac{t}{\Delta r_{0}} \label{eqn:delta_r}
\end{equation}
where $d^2 \omega(k_{0})/dk^{2}$ is the rate of change of the group velocity of the pulse with respect to $k$. Eqn.~\ref{eqn:delta_r} can then be fitted to the FWHM measurements, allowing $d^2 \omega(k_{0})/dk^{2}$ to be determined for each passband.

The expansion rate and the resulting value of $d^2 \omega(k_{0})/dk^{2}$ from the bottom-left panel of Figure~\ref{fig:kinematics} are given in Table~\ref{tbl:kinematics} for each passband. The general expansion rate in each case is positive within error, indicating statistically significant pulse broadening. This implies that CBFs are dispersive pulses, confirming the results of \citet{Warmuth:2004ab,Veronig:2010ab} and \citet{Long:2011ab}.

\subsection{Temperature Dependence}
\label{subsect:discrepancies}

The kinematics of the CBF could be derived for individual passbands at different peak emission temperatures due to the very high cadence of \emph{SDO} (see Table~\ref{tbl:kinematics}). A spread is apparent in both the initial velocity and acceleration of the pulse, from $\sim$380 to $\sim$460~km~s$^{-1}$ and $\sim -128$ to $\sim -430$~m~s$^{-2}$ respectively. This variation was studied by making a comparison with the peak emission temperatures ($T_{peak}$) of the different AIA passbands \citep[as given in Table~\ref{tbl:kinematics} and discussed by][]{odwyer2010}. 

It was found that the CBF kinematics and $T_{peak}$ for each passband are inversely related. As temperature tends to increase while density and magnetic field strength decrease with height in the quiet Sun, this implies that in cooler, more dense plasma the CBF has a higher velocity. This is characteristic of a compressive pulse and combined with the dispersion and deceleration indicates that the CBF is best described as a magnetohydrodynamic wave pulse. The randomly structured nature of the quiet corona suggests that any globally-propagating pulse must traverse magnetic field lines, indicating a fast-mode rather than slow-mode CBF interpretation.

The CBF morphology across different passbands shows some discrepancies that invite further investigation, particularly the simultaneous intensity decrease at 171~\AA\ and increase in the cooler 304~\AA\ passband. The 171~\AA\ emission drop (visually identified here but not tracked) has been characterised as evidence of plasma heating from 171~\AA\ into the 193, 211, and 335~\AA\ passbands \citep{Wills-Davey:1999ab,Liu:2010ab}. This heating implies that the CBF pulse is coronal, an observation consistent with the height measurements made by \citet{kienreich2009} and \citet{patsourakos2009}, but complicated by the increase in 304~\AA\ emission (dominated by two chromospheric \ion{He}{2} lines at 303.781 and 303.786~\AA).  Although there is also a coronal \ion{Si}{11} emission line at 303.33~\AA, \citet{odwyer2010} have noted that this line does not make a notable contribution to AIA quiet Sun observations, suggesting that the observed intensity increase must be due to \ion{He}{2} emission. 

The formation mechanism of \ion{He}{2} emission has been the subject of detailed investigation \citep[see e.g.,][]{MacPherson:1999ab,Andretta:2003ab,Jordan:2007ab} due to its complex nature, with results suggesting that it is formed by collisional excitation from thermal electrons in the quiet corona. The increased temperature gradient caused by passage of a compressive coronal pulse could enhance this effect, producing the observed 304~\AA\ intensity increase. The CBF would therefore be coronal, as predicted by the observed drop in 171~\AA\ intensity.

\subsection{Coronal Seismology}
\label{subsect:seismology}

The passband-dependent kinematics indicate that the pulse morphology is significantly influenced by the plasma through which it propagates. By examining how the plasma affects the kinematics for each passband, it is possible to directly quantify the characteristics of the quiet coronal plasma. For example, the fast-mode wave speed is defined as,
\begin{equation}\label{eqn:v_fast_mode}
v_{fm} = \sqrt{v_A^2 + c_s^2},
\end{equation}
where the Alfv\'{e}n speed and sound speed are $v_A = B/(4 \pi n m)^{1/2}$ and $c_s = (\gamma k T/m)^{1/2}$ respectively. Here $B$ is the magnetic field strength, $n$ is the particle density, $m$ is the proton mass, $\gamma$ is the adiabatic index (typically $5/3$), $k$ is Boltzmann's constant and $T$ is the peak emission temperature ($T_{peak}$; the values used are given in Table~\ref{tbl:kinematics}). If the CBF pulse is treated as a fast-mode wave then the final pulse velocity (i.e.,\ the velocity of the pulse when it can no longer be detected by the algorithm) must be the fast-mode velocity of the given passband, since the pulse can not propagate below this velocity. These values are given in Table~\ref{tbl:kinematics} for each \emph{SDO} passband studied.

By taking the peak emission temperature of each passband as the temperature, only the magnetic field strength and density are unknown in the above equations. Coronal magnetic field strength estimates typically involve extrapolating photospheric magnetic field measurements into the corona and are not very well constrained (particularly in the quiet Sun). In contrast, coronal densities can be estimated using density sensitive line ratios \citep{Gallagher:1999ab} and are well-constrained.

The above equations can be rearranged to give,
\begin{equation}\label{eqn:b_n_variation}
B = \sqrt{4 \pi n (m v^2_{fm} - \gamma k T_{peak})},
\end{equation}
implying that the quiet coronal magnetic field strength may be estimated using the derived CBF kinematics. The final velocity values given in Table~\ref{tbl:kinematics} were combined with a range of typical quiet coronal densities \citep[$\sim$2--6$\times10^{8}$~cm$^{-3}$; see][for more details]{Wills-Davey:2007oa} to produce an estimated quiet coronal magnetic field strength range of $\sim$1--2~G. This is comparable to the value derived by \citet{West:2011ab} from detailed \emph{STEREO}/EUVI kinematic estimates and \emph{Hinode}/EIS density measurements. The good agreement of the range estimated here with the work of \citet{West:2011ab} indicates that our assumptions are correct and CBFs can be used to probe the physical characteristics of the plasma through which they propagate.

\section{Discussion and Conclusions}
\label{sect:disc_and_conc}

Comparing EUVI and high cadence AIA observations of the 2010~August~14 CBF event allowed an examination of the accuracy of previous CBF kinematics estimates, which involved combining distance-time measurements from different passbands due to a paucity of data \citep[e.g., ][]{Long:2008eu,Patsourakos:2009ab,kienreich2009,Veronig:2010ab}. While this was necessary to derive kinematics from the small data-sets available, our results indicate that this approach underestimated the general kinematics of the CBF. It may have also masked the pulse acceleration and did not detail the effect of the plasma on the pulse. The presence of deceleration in both EUVI and AIA data (despite the different cadence and spacecraft positions), also suggests that it is characteristic of the phenomenon.

The clear dispersion apparent in both EUVI and AIA data confirms the observations of \citet{Warmuth:2004ab,Veronig:2010ab} and \citet{Long:2011ab}. These repeated measurements of significant pulse broadening strongly indicate that CBFs have a dispersive nature which, allied to the traditional point-and-click techniques for identifying them, may have contributed to the uncertainty surrounding their acceleration. When both the dispersion and deceleration are considered, CBFs may be best described using a wave interpretation. Although this behavior is not predicted by ideal MHD wave theory, it is consistent with the results of \citet{Murawski:2001ab} and \citet{Nakariakov:2005ab}, and the randomly structured nature of the corona. The dispersion relation of the pulse was determined by treating it as a linear superposition of sinusoidal waves within a Gaussian envelope, allowing an insight into its physical nature. 

The CBF pulse was observed to display kinematics that were dependent on the passband studied; a unique result that supports the wave interpretation of CBFs. In particular, the pulse exhibited a compressive nature, appearing to propagate at a higher velocity with stronger deceleration in cooler, denser plasma. This is the first observation of this property of CBFs, and is a result of the very high cadence capabilities of \emph{SDO}/AIA. This kinematic variation also provides a simple diagnostic of the emitting plasma in each passband, allowing coronal seismology to be used to determine the physical parameters of the corona directly.

Even though CBF propagation has previously been proposed as a way of directly probing the structure of the solar corona, this has been complicated by their uncertain physical nature. The results presented here, in addition to recent work by \citet{Patsourakos:2010ab,kienreich2011} and \citet{Long:2011ab} indicate that CBFs are fast-mode MHD waves, allowing them to be used to examine the environment through which they propagate. Alternative techniques can be used to determine typical densities and the temperature of the different passbands, allowing the magnetic field strength to be estimated using the CBF. The range of values derived here ($\sim$1--2~G) are comparable to those estimated by \citet{West:2011ab} \citep[and typically assumed for the quiet corona, e.g.,][]{Wills-Davey:2007oa}, indicating that CBFs can be used to directly probe the plasma through which they propagate. 

These results are most compatible with the wave interpretation of a CBF pulse. The observed dispersion implies that CBFs are not accurately described by the soliton model proposed by \citet{Wills-Davey:2007oa}, while the CBF height range (on-disk near the limb over an extended time period in both \emph{SDO} and \emph{STEREO} observations) is inconsistent with the progressively higher emission predicted by \citet{Delannee:2008uq}. The multi-temperature emission does not match the low foot-point signature predicted by \citet{Attrill:2007vn} and there was no indication of the additional coronal Moreton wave predicted by \citet{Chen:2002rw}. Although the initial driver is uncertain, the CBF could be a product of the rapid over-expansion of the erupting CME bubble \citep[cf.][]{Patsourakos:2010bc} before decoupling and propagating freely. The high cadence observations available from \emph{SDO} will allow this issue to be resolved.

\acknowledgments
D.M.L. is a 2010-2011 Pre-Doctoral Fellow at the Harvard-Smithsonian Center for Astrophysics, and carried out some of this work while a Government of Ireland Scholar supported by the Irish Research Council for Science, Engineering and Technology (IRCSET). We wish to thank S. Patsourakos, R.~T.~J. McAteer, D.~S. Bloomfield and J. Raymond for useful discussions, and the anonymous referee whose comments helped to improve this paper.

\clearpage

\clearpage

\clearpage

\begin{deluxetable}{cccccccc}
\tablecolumns{7}
\tabletypesize{\footnotesize}
\tablewidth{0pt}
\centering
\tablecaption{2010~August~14 CBF properties\label{tbl:kinematics}}
\tablehead{
\colhead{Spacecraft} & \colhead{Passband} & \colhead{$T_{peak}$\tablenotemark{a}} & \colhead{$v_{0}$} & \colhead{$a_{0}$} & \colhead{Expansion Rate} & \colhead{$d^2 \omega(k_{0})/dk^{2}$} & \colhead{$v_{\mathrm{final}}$} \\ 
\colhead{} & \colhead{\AA} & \colhead{MK} & \colhead{km~s$^{-1}$} & \colhead{m~s$^{-2}$} & \colhead{km~s$^{-1}$} & \colhead{Mm$^{2}$~s$^{-1}$} & \colhead{km~s$^{-1}$}
}
\startdata
{\it STEREO}-A 	& 195 					& 1.5		& $343\pm52$	& $-71\pm69$ 	& $130.6\pm12.3$ 	& $20.80\pm2.08$	& \nodata \\ 
{\it SDO} 		& All\tablenotemark{b} 	& \nodata	& $411\pm17$ 	& $-279\pm36$ 	& $222.0\pm1.8$ 	& $10.38\pm0.20$	& \nodata \\
 				& 335 					& 2.5		& $379\pm12$ 	& $-128\pm28$ 	& $211.5\pm4.9$ 	& $13.32\pm0.53$	& $273\pm35$ \\ 
 				& 211 					& 1.8		& $409\pm11$ 	& $-298\pm24$ 	& $238.1\pm2.3$ 	& $8.37\pm0.27$		& $144\pm32$ \\ 
 				& 193 					& 1.6		& $419\pm5$ 	& $-318\pm13$ 	& $190.4\pm5.2$	 	& $11.05\pm0.55$	& $163\pm15$ \\ 
 				& 304 					& 0.05		& $460\pm28$ 	& $-431\pm86$ 	& $214.4\pm7.1$ 	& $13.17\pm0.68$	& $181\pm84$ \\ 
\enddata
\tablenotetext{a}{$T_{peak}$ here refers to the peak emission temperature of each passband.}
\tablenotetext{b}{Distance-time measurements for all passbands observed by AIA were combined for comparison.}
\end{deluxetable}

\end{document}